\documentclass[prl, twocolumn]{revtex4}
\usepackage{graphicx}

\begin{document}
\newcommand{\e}{\mathrm{e}}
\newcommand{\x}{{\bf r}}
\newcommand{\K}{{\bf k}}
\newcommand{\y}{{\bf y}}
\newcommand{\D}{{\rm d}}
\newcommand{\p}{{\bf p}}
\newcommand{\tr}{\mathrm {Tr}}
\newcommand{\de}{:=}

\title{Quantum multimode model of elastic scattering from Bose Einstein condensates}

\author{P. Zi\'{n}$^1$, J. Chwede\'{n}czuk$^1$, A. Perez$^1$,
K. Rz\c{a}\.zewski$^2$ and M. Trippenbach$^1$}

\affiliation{$^1$ Physics Department, Warsaw University, Ho\.{z}a 69, PL-00-681 Warsaw, Poland,
$^2$ Center for Theoretical Physics, Polish Academy of Science, Al. Lotnik\'{o}w 32/46, PL-02-668
Warsaw, Poland.}

\begin{abstract}
Mean field approximation treats only coherent aspects of the evolution of a Bose Einstein
condensate. However, in many experiments some atoms scatter out of the condensate. We study an
analytic model of two counter-propagating atomic Gaussian wavepackets incorporating dynamics of
incoherent scattering processes. Within the model we can treat processes of elastic collision of
atoms into the initially empty modes, and observe how, with growing occupation, the bosonic
enhancement is slowly kicking in. A condition for bosonic enhancement effect is found in terms of
relevant parameters. Scattered atoms form a squeezed state that can be viewed as a multi-component
condensate.  Not only are we able to calculate the dynamics of mode occupation, but also the full
statistics of scattered atoms.
\end{abstract}

\maketitle

A remarkably universal tool describing vast majority of experiments with the Bose Einstein
condensates is the celebrated Gross-Pitaevskii equation [GPE]. It describes a coherent evolution
of the atomic mean field. In the Hartree interpretation, its time-dependent version assumes that
each atom of the system undergoes identical evolution. This is a good assumption since in typical
experiments the wave-packet of the system contains many thousands of particles in the same state.
To use a term borrowed from quantum optics, the time-dependent GPE describes stimulated processes.
In some experiments~\cite{kozuma}, however,  there is a clear evidence of spontaneous
processes. For example, in a collision between two condensates, some atoms from colliding quantum
matter droplets would inevitably scatter away from them. This is a loss process, which is not
accounted for by the conventional GPE. Description of such phenomena calls for use of quantum
fields instead of c-number wave-functions. This is not easy since, in general, field equations are
nonlinear. Instead of quantum fields, several groups used classical stochastic fields to imitate
quantum initiation of spontaneous processes ~\cite{Chwed}. At this point it is hard to access the
accuracy of these methods. Solid results so far has only been obtained within perturbation theory
~\cite{Bach, Band, Yuro}. It is the purpose of this Letter to present the first exact
nonperturbative calculation of collisional losses, valid in the regime of Bose enhancement. Our
model assumes spherical nonspreading  Gaussians for the colliding wave-packets. No doubt it will
serve as a benchmark test of validity of various approximate schemes including classical
stochastic fields.

A system of Bosons interacting via contact potential is described by the Hamiltonian
\begin{eqnarray}
\label{ham1}\mathrm{\hat H}&=&-\int\D^3 r \, \hat \Psi^\dagger({\bf
r},t)\frac{\hbar^2\nabla^2}{2m}\hat\Psi({\bf r},t)
\nonumber\\
&+&\frac{g}{2}\int \mbox{d}^3 r \, \hat{\Psi}^\dagger({\bf r},t) \hat{\Psi}^\dagger({\bf r},t)
\hat{\Psi}({\bf r},t) \hat{\Psi}({\bf r},t),
\end{eqnarray}
where $\hat\Psi(\x)$ is a field operator satisfying equal time bosonic commutation relations, $m$
is the atomic mass and $g$ determines the strength of the inter-atomic interactions. Since the
Hamiltonian (\ref{ham1}) is of the fourth order in $\hat\Psi$, the Heisenberg equation governing
the evolution of the field will be nonlinear and thus, in general, analytically and numerically
untractable. However, for some physical systems, a Bogoliubov approximation can be applied leading
to linear Heisenberg equations. The idea underlying this approximation states that for some cases
the field operator might be split into two parts $\psi$ and $\hat\delta$. First contribution
describes macroscopically occupied field and since its fluctuations are usually small, its
operator character might be dropped ($\psi$ becomes a c-number wave-function satisfying GPE). The
second part $\hat\delta$, representing fluctuations, will require full quantum mechanical
treatment, but as long as we neglect its back-reaction on $\psi$ the evolution of $\hat\delta$
will be linear.

In this Letter we consider a process of collision of two strongly occupied Bose Einstein
condensates. Initial state of the system consists of two counter-propagating atomic wave-packets
and the ``sea'' of unoccupied modes. For such a system the Bogoliubov approximation can be
applied. The splitting of the bosonic field is performed in the following manner:
\begin{eqnarray}
\hat\Psi(\x, t)&=&\psi_Q(\x, t)+\psi_{-Q}(\x, t)+\hat{\delta}(\x, t),
\label{deco1}
\end{eqnarray}
where the subscript $\pm Q$ denotes the mean momentum of the colliding condensates. Upon inserting
Eq.~(\ref{deco1}) into the Hamiltonian (\ref{ham1}) one obtains a collection of different terms.
We keep only those, that lead to creation or annihilation of a pair of particles
\begin{eqnarray}
\label{ham2} \mathrm{H}&=&-\int \mbox{d}^3 r \, \hat{\delta}^
\dagger({\bf r},t)\frac{\hbar^2\nabla^2}{2m}\hat{\delta}({\bf r},t)
\\ \nonumber
&+&g\int \mbox{d}^3 r \, \hat{\delta}^\dagger({\bf r},t)
\hat{\delta}^\dagger({\bf r},t) \psi_Q({\bf r},t) \psi_{-Q}({\bf r},t)+
\mathrm{H.c.}
\end{eqnarray}
One can argue that such an approximation gives correct results if and only if the kinetic energy
associated with the center-of-mass motion is much larger than the interaction energy per particle,
$\hbar^2Q^2/(2m) \gg gn$, where n is the average density of the particles in the condensates.
Numerical proof of the above statement in the simplest case of two plane matter waves was given in
\cite{Bach}. This condition is readily fulfilled in current
experiments~\cite{kozuma,stenger,chikka} and all the results below are obtained in this regime.

In order to further simplify the dynamics we compare three characteristic timescales that appear
in the problem; the collisional time, $t_{C}=(m\sigma)/(\hbar Q)$,  the time it takes for each
wave-packet to pass through its colliding partner, the linear dispersion time,
$t_{LD}=m\sigma^2/\hbar$ \cite{Tripp}, characteristic time of the spread of the wave-packet due to
kinetic energy term (neglecting the nonlinearity), and nonlinear dispersion time,
$t_{ND}=\sqrt{\pi^{3/2} m\sigma^5/gN}$, time of ballistic expansion in Thomas Fermi approximation
\cite{CDG}.
Here each of the wave-packets has the radius of $\sigma$ and contains $N/2$ atoms. The dynamics of
our system depends on the relations between timescales defined above. Hence we introduce
dimensionless parameters: $t_{LD}/t_C=\beta$ and $(t_{LD}/t_{ND})^2=\alpha$. When the number of
elastically scattered atoms is small in comparison with the total number of atoms in both
wave-packets and both linear and nonlinear dispersion timescales are much longer than the
collisional time ($(t_{LD}/t_C)=\beta \gg 1$ and $(t_{ND}/t_C)^2=\beta^2/\alpha\gg 1$), we can
neglect the change of shape of the macroscopically occupied functions $\psi_Q({\bf r},t)$ during
the collision. In our model we use spherically symmetric Gaussian wave-functions
\begin{eqnarray}
\label{cond} &&\psi_{\pm Q}({\bf r},t) = \sqrt{\frac{N}{2\pi^{3/2}\sigma^{3}}}  \exp \left[ \pm
iQx_1
-\frac{i\hbar t Q^2}{2m} \right] \times\nonumber\\
&&\times\exp\left[-\frac{1}{2\sigma^2}\left(\left(x_1\mp\frac{\hbar
Qt}{m}\right)^2+x_2^2+x_3^2\right)\right],
\end{eqnarray} where ${\bf r}=(x_1,x_2,x_3)$. In the dimensionless units, ($t\equiv t/t_C $ and
$x_i \equiv x_i/\sigma$, for $i=1,2,3$), the Heisenberg evolution equation of the field operator
$\hat\delta\equiv \hat\delta \exp \left(i \beta t/2\right)$ can be obtained upon substituting
(\ref{cond}) into (\ref{ham2})
\begin{equation}
i\beta\partial_t\hat{\delta}(\x,t)=-\frac{1}{2}\left(\Delta+\beta^2\right) \hat{\delta}(\x,t)
+\alpha e^{-r^2-t^2}\hat{\delta}^\dagger(\x,t).\label{EofM}
\end{equation}
The above equation has spherical symmetry! Hence, we decompose $\hat{\delta}$ into the basis of
spherical harmonics
\begin{eqnarray}
\label{decomp}\hat{\delta}(\x,t)&=&\sum_{n,l,m}R_{n,l}(r)Y_{lm}(\theta,\phi)\hat a_{n,l,m}(t),
\end{eqnarray}
where $\hat a_{n,l,m}$ are annihilation operators for a particle in the mode described by $n,l,m$
quantum numbers. There is still a freedom of choice with regards to the set of orthogonal
functions $R_{n,l}(r)$. As we shall see below a good candidate is a set of eigenfunctions of
spherically symmetric harmonic oscillator,
\begin{equation}
R_{n,l}(r) = \sqrt{\frac{2n!a_0^{-3}}{\Gamma(l+n+\frac{3}{2})}} \left( \frac{r}{a_0}\right)^l
e^{-\frac{r^2}{2a_0^2}}L_n^{l+\frac{1}{2}}\left(\frac{r^2}{a_0^2}\right),
\end{equation}
where $L_n^{l+\frac{1}{2}}(x)$ is the associated Laguerre polynomial \cite{Abram} and $a_0$, a
harmonic oscillator length, is an auxiliary free parameter that can be chosen to minimize the
computational effort.
The evolution of $\hat a_{n,l,m}(t)$ is described by
\begin{eqnarray}
&&\label{evolution} i\partial_t\hat a_{n,l,m}=\frac{E_{n,l}-\beta^2}{2\beta} \hat
a_{n,l,m}+D_{n,l}
\hat a_{n-1,l,m}+\nonumber\\
&&D_{n+1,l}\hat a_{n+1,l,m} +\frac{\alpha}{\beta}\e^{-t^2}\sum_{n'}C_{n,n',l} \hat
a^\dagger_{n',l,-m}, \label{lsystem}
\end{eqnarray}
where coefficients $D_{n,l}= \sqrt{n(n+l+1/2)}/(2\beta a_0^2)$, $E_{n,l}=(2n+l+3/2)/a_0^2$ and
\begin{eqnarray} \nonumber
&&C_{n,n',l} = \int_0^\infty r^2 \mbox{d}r \, R_{n,l}(r) \exp(-r^2) R_{n'l}(r)= \\ \nonumber \\
\nonumber &&=\sqrt{\frac{\Gamma \left(n+l +\frac{3}{2}\right)\Gamma \left(n'
+l+\frac{3}{2}\right)}{\Gamma \left(l+\frac{3}{2} \right)^2\Gamma
\left(n+1\right)\Gamma\left(n'+1\right)}}
\left(1+ a_0^2\right)^{-l-\frac{3}{2}}\times\nonumber\\
&&\times \left[\frac{ a_0^2}{1+ a_0^2}\right]^{n+n'} F \left( -n,-n',l+\frac{3}{2} , 1/a_0^4
\right).
\end{eqnarray}
Here $F(a,b,c,x)$ is a hypergeometric function \cite{Abram}. Notice that all coupling coefficients
are calculated analytically and the $\hat a_{n,l,m}$ operators for different $l$ and $m$ are
decoupled. Moreover, equations (\ref{evolution}) do not depend on quantum number $m$. With all
these simplifications the linear system of equations (\ref{lsystem}) can be solved numerically.



\begin{figure}[htb]
\centering
\includegraphics[scale=0.3, angle=270]{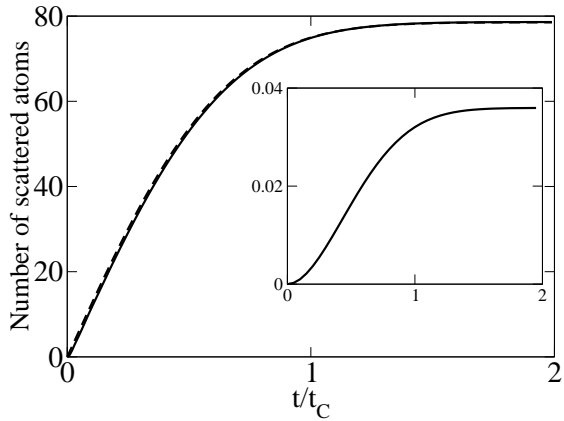}
\caption{Number of scattered atoms versus time in perturbative regime; dashed line - analytical
result given by (\ref{erf}). Solid line - numerical result obtained from our model (using
Eq.~(\ref{number})) for $\alpha=20$ and $\beta=60$. The inset shows the time evolution of the
largest eigenvalue of the density matrix.} \label{fig1}
\end{figure}

The solution of the set of dynamical equations \ref{evolution} for $\hat a_{n,l,m}$ contains the
full information about the considered quantum system. In particular, we can reconstruct the
operator $\hat\delta(\x,t)$, using decomposition defined in Eq.~(\ref{decomp}). The most
straightforward observable quantity, the number of elastically scattered atoms as a function of
time can be expressed in terms of the trace of the density matrix
\begin{eqnarray}
\label{number}\mathcal{S}(t)&=&\int d^3 r
\langle\hat\delta^\dagger(\x,t)\hat\delta(\x,t)\rangle= \nonumber
\\&&=
\sum_{n=0}^{\infty}\sum_{l=0}^{\infty}\left(2l+1\right)\langle
\hat a^\dagger_{n,l,m}(t)\hat a_{n,l,m}(t)\rangle,
\end{eqnarray}
where $(2l+1)$ accounts for the degeneracy of
Eq.~(\ref{evolution}) with regards to the quantum number
$m$~\cite{remark3}.
In the limit where $\alpha/\beta$ is small (notice that in Eq.~(\ref{evolution}), this coefficient
multiplies the source term), $S(t)$ can be evaluated in the first order perturbation approximation
giving~\cite{Bach}
\begin{equation}
\mathcal{S}(t)=\frac{\pi\alpha^2}{16}\,\mathrm{erf}\left(\sqrt2t\right).\label{erf}
\end{equation}
The same result is obtained using imaginary scattering length method ~\cite{Band}. Quality of this
approximation is illustrated in Fig.\ref{fig1}.

The bonus of having solved the full set of operator equations is that calculating full density
matrix of the system of scattered atoms
($\rho(\x,\x',t)=\langle\hat\delta^\dagger(\x,t)\hat\delta(\x',t)\rangle$) or even higher order
correlation functions is just as easy as finding $\mathcal{S}(t)$. In the basis (\ref{decomp}),
due to the decoupling property,  density matrix can be written as a direct product of
$\rho_{n,n',l,m}=\langle \hat a^\dagger_{n,l,m}(t)\hat a_{n',l,m}(t)\rangle$ matrices, for
different $l$ and $m$.
In the inset of Fig.\ref{fig1} we present the time evolution of the largest of the eigenvalues of
the density matrix $\rho(\x,\x',t)$. Due to the normalization of the density matrix,
$\sum_i\lambda_i(t)=\mathcal{S}(t)$, where $\lambda_i$ are the eigenvalues of the density matrix,
the inset of Fig.\ref{fig1} shows that for $\alpha=20$, $\beta=60$ there is much less than one
particle even in the mostly populated eigenmode.

\begin{figure}[htb]
\centering
\includegraphics[scale=0.30, angle=270]{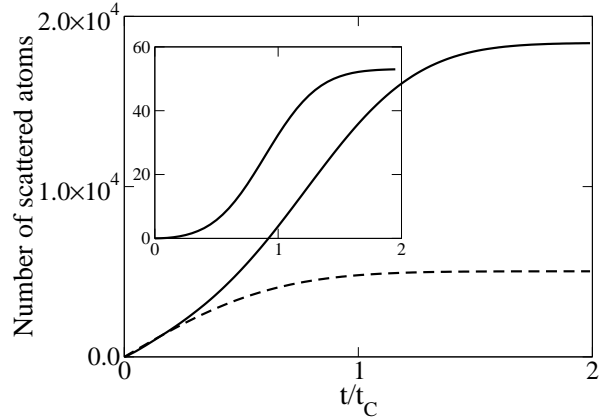}
\caption{Number of scattered atoms versus time in non-perturbative regime where the bosonic
enhancement occurs. Dashed line -- analytical result given by (\ref{erf}). Solid line -- numerical
result obtained from (\ref{number}). Parameters are: $\alpha=160$, $\beta=40$. The inset shows the
time evolution of the biggest eigenvalue of the density matrix.} \label{fig2}
\end{figure}

Figure \ref{fig2} shows analogous comparison between  perturbative solution (\ref{erf}) and
formula (\ref{number}) in the regime of parameters where the perturbation theory is expected to
fail (the criterion for bosonic enhancement is  $\alpha/\beta > 1$ \cite{future}). The figure
shows that until some critical time, approximately equal to $0.2\, t_C$, both the perturbative and
full solutions agree very well. At this critical time the formula (\ref{number}) exceeds the
perturbative solution and the difference between curves rapidly grows in time. At the same time
the biggest eigenvalue of the density matrix of the system reaches one, which means that there is
one particle in the mostly populated eigenmode. This observation gives explanation to the growing
discrepancy between two curves shown in Fig.\ref{fig2}. Once approximately one atom is scattered
into one of the eigenmodes of the density matrix the probability of scattering another atom into
this mode grows rapidly. This is due to bosonic statistics of the scattered atoms and is called
bosonic enhancement effect.

\begin{figure}[htb]
\centering
\includegraphics[scale=0.3,angle=270]{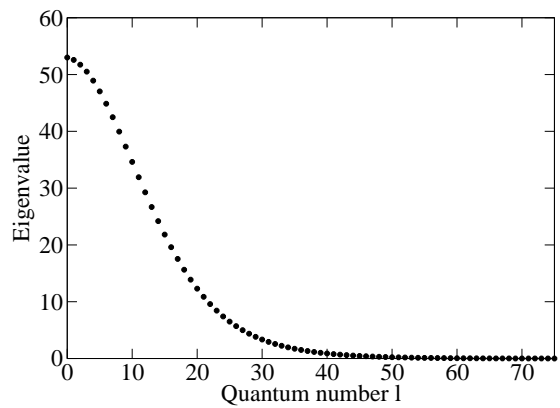}
\caption{The biggest eigenvalue of the density matrix for different $l$ for $\alpha=160$ and
$\beta=40$, at time $t=2t_C$. Several eigenvalues of the same order indicate the presence of the
quasi-condensate.} \label{fig5}
\end{figure}

An interesting information about the system might be obtained upon analyzing the largest
eigenvalues of $\rho_{n,n',l,m}$ for each quantum number $l$. Figure \ref{fig5} juxtaposes these
eigenvalues as a function of $l$, for the case with bosonic enhancement. The plot shows that the
density matrix has several eigenvalues of the same order. Such a system is similar to
quasi-condensate, in contrast to the commonly used definition of the condensate as described by a
density matrix having one dominant eigenvalue~\cite{Zagrenbov}.

\begin{figure}[htb]
\centering
\includegraphics[scale=0.30,angle=270]{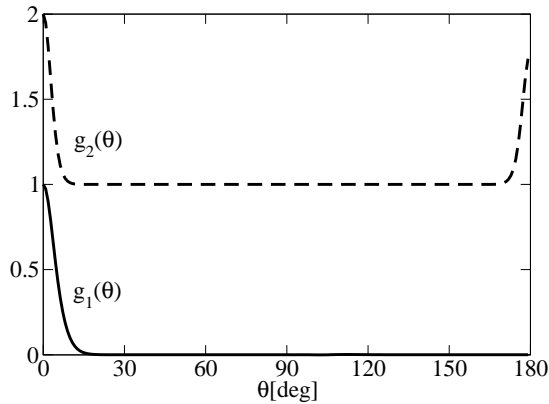}
\caption{First and second order correlation functions in  momentum space $g_1(\K, \K')$ and
$g_2(\K, \K')$ for $|\K|=|\K'|=Q$, as a function of relative azimuthal angle $\theta$ at $t=2t_C$
for $\alpha=160$ and $\beta=40$.} \label{fig4}
\end{figure}

From the experimental point of view, coherence properties  of the scattered atoms are of great
importance. These properties are best characterized by the correlation functions. In particular,
the first and second order correlation functions can be measured in experiment. In one of the most
commonly used method, time--of--flight measurement, the momentum distribution of the system is
obtained. Thus here we calculate the first and second order correlation functions in momentum
space using
\begin{eqnarray}
\label{g1}g_1(\K, \K', t)&=&\frac{\langle\hat\delta^\dagger(\K, t)\hat\delta(\K',
t)\rangle}{\sqrt{\langle\hat\delta^\dagger(\K, t)\hat\delta(\K,
t)\rangle\langle\hat\delta^\dagger(\K', t)\hat\delta(\K', t)\rangle}}
\end{eqnarray}
for the former, and
\begin{eqnarray}
\label{g2}g_2(\K, \K', t)&=&\frac{\langle\hat\delta^\dagger(\K, t)\hat\delta^\dagger(\K',
t)\hat\delta(\K', t)\hat\delta(\K, t)\rangle}{\langle\hat\delta^\dagger(\K, t)\hat\delta(\K,
t)\rangle\langle\hat\delta^\dagger(\K', t)\hat\delta(\K', t)\rangle}
\end{eqnarray}
for the latter. Due to spherical symmetry of Heisenberg equation for $\hat\delta$, the momentum
density $\langle\hat\delta^\dagger(\K, t)\hat\delta(\K, t)\rangle$ is spherically symmetric as
well. Moreover, since the Hamiltonian (\ref{ham2}) is quadratic in $\hat\delta$ and the initial
state is a vacuum state, than, in Schr\"{o}dinger picture, at any later time $t$ the state of
scattered atoms is a multimode squeezed state~\cite{xin}. According to general properties of
multimode squeezed states, the $n$-th order correlation function $g_n(\K, \K')$ for $\K=\K'$ is
equal to $n!$. It is confirmed by our numerical results.
The solid line in Fig.\ref{fig4} shows the first order correlation function (\ref{g1}) plotted for
fixed length of the $\K$ and $\K'$ vectors ($|\K|=|\K'|=Q$) as a function of relative angle
$\theta$. As expected, for $\theta=0$ the condition, $g_1(\K, \K)=1$ is satisfied. Also, the
limited coherence angle, due to spontaneous initiation of scattering process is clearly visible.
The dashed line Fig.\ref{fig4} shows the second order correlation function (\ref{g2}). Once again,
a prediction $g_2(\K, \K)=2$ is met. As Fig.\ref{fig4} shows, the $g_2$ function reveals strong
correlation between atoms scattered in direction $\K$ and $-\K$ which corresponds to relative
angle $\theta=180^\circ$. This is an intuitive result, since atoms get scattered in pairs in such
a way that the momentum and energy conservation laws are satisfied. Finally, the width of the
correlation peak of $g_2$ in the forward direction in the perturbative regime scales as $1/\beta$,
which is proportional to the size of colliding wave-packets~\cite{future}. This is in analogy to
Hanburry-Brown and Twiss method of estimating sizes of distant stars by measuring
intensity-intensity correlation function~\cite{Hanburry} and relating density-density correlation
of $\pi$-mesons to the size of fireball in high energy collision of hadrons~\cite{Kopylov}.

In conclusion, upon analyzing the quantum model of two counter-propagating atomic Gaussian
wave-packets we get a deeper insight into processes of elastic collision losses of atoms and are
able to study the transition from spontaneous regime to the bosonic enhancement.  Scattered atoms
form a squeezed state that can be viewed as a multi-component condensate. Within this model in
principle all order correlation functions are accessible and hence it has a high predictive power.

The authors acknowledge support from KBN Grant 2P03 B4325 (J. Ch., P.Z.), Polish Ministry of
Scientific Research and Information Technology under grant PBZ-MIN-008/P03/2003 (M. T., K.R.).


\begin{thebibliography}{30}

%
%
%

\bibitem{kozuma} M. Kozuma {\it et al.}, Phys. Rev. Lett. {\bf 82}, 871 (1999); J. M. Vogels, K. Xu, and W. Ketterle,  Phys. Rev. Lett. {\bf 89}, 020401 (2002)

\bibitem{Chwed} J. Chwede\'{n}czuk, M. Trippenbach and K. Rz\c{a}\.{z}ewski, J. Phys. B, L391, {\bf 37}
(2004); A.A. Norrie, R.J. Ballagh and  C.W. Gardiner, cond-mat/0403378

\bibitem{Bach} R. Bach, M. Trippenbach and K. Rz\c{a}\.zewski, Phys. Rev. A {\bf 65} 063605 (2002)

\bibitem{Band} Y. B. Band {\it et al.}, Phys. Rev. Lett. {\bf 84}, 5462 (2000)

\bibitem{Yuro} V. A. Yurovsky, Phys. Rev. A {\bf 65} 033605 (2002)

\bibitem{stenger} J. Stenger {\it et al.}, Phys. Rev. Lett. {\bf 82}, 4569 (1999).

\bibitem{chikka}A. P.Chikkatur {\it et al.}, Phys. Rev. Lett. {\bf 85}, 483 (2000).

\bibitem{Tripp}
M. Trippenbach, Y. B. Band, and P. S. Julienne,  Phys. Rev. A {\bf 62}, 023608 (2000)

\bibitem{CDG}
Y. Castin and R. Dum,  Phys. Rev. Lett. {\bf 77}, 5315-5319 (1996);
Yu. Kagan, E. L. Surkov, and G. V. Shlyapnikov,  Phys. Rev. A {\bf 55}, R18-R21 (1997)

\bibitem{Abram} M. Abramovich, I. A. Stegun, "Handbook of Mathematical
Functions With Formulas, Graphs and Mathematical Tables",  Dover Publications 1974.

\bibitem{Zagrenbov} V.A. Zagrebnov, J.-B. Bru, Physics Reports {\bf 350}, 291-434(2001)

\bibitem{future} J. Chwedeñczuk {\it et al.}, in preparation.

\bibitem{Hanburry}  R. Hanburry-Brown and R. Q. Twiss, Phil. Mag. {\bf 45}, 633 (1954)

\bibitem{Kopylov} G. I. Kopylov, Phys. Lett. {\bf 50B}, 472 (1974)

\bibitem{Keith} T. Kohler and K. Burnett, Phys. Rev A {\bf 65}, 033601 (2002)


\bibitem{Deng} L. Deng {\it et al},  Nature (London) {\bf 398}, 218 (1999)




\bibitem{remark3}  The expectation value in
equation (\ref{number}) and below it are taken in the initial vacuum state $|0\rangle$, satisfying
the condition $\hat a_{n,l,m}(0)|0\rangle=0$ for all $n$, $l$ and $m$.

\bibitem{xin} X. Ma and W. Rhodes, Phys. Rev A {\bf 41}, 4625 (1990)

\end{thebibliography}
\end{document}